\documentclass[twocolumn,english,aps,english,showpacs,preprintnumbers,groupedaddress,amsmath,amssymb,pra]{revtex4}
\usepackage[T1]{fontenc}
\usepackage[latin9]{inputenc}
\usepackage{amsmath}
\usepackage{amssymb}
\usepackage{graphicx}
\usepackage{esint}

\makeatletter

\newcommand{\noun}[1]{\textsc{#1}}

\@ifundefined{textcolor}{}
{%
 \definecolor{BLACK}{gray}{0}
 \definecolor{WHITE}{gray}{1}
 \definecolor{RED}{rgb}{1,0,0}
 \definecolor{GREEN}{rgb}{0,1,0}
 \definecolor{BLUE}{rgb}{0,0,1}
 \definecolor{CYAN}{cmyk}{1,0,0,0}
 \definecolor{MAGENTA}{cmyk}{0,1,0,0}
 \definecolor{YELLOW}{cmyk}{0,0,1,0}
}

\usepackage{babel}

\makeatother

\usepackage{babel}
\begin{document}

\title{Slow light in saturable absorbers: Progress in the resolution of a controversy}

\author{Bruno Macke}

\author{Igor Razdobreev}

\author{Bernard Ségard}
\email{bernard.segard@univ-lille1.fr}

\affiliation{Laboratoire de Physique des Lasers, Atomes et Molécules , CNRS et
Université de Lille, F-59655 Villeneuve d'Ascq, France}

\date{20 June 2017}
\begin{abstract}
There are two opposing models in the analysis of the slow transmission
of light pulses through saturable absorbers. The canonical incoherent
bleaching model simply explains the slow transmission by combined
effects of saturation and of non-instantaneous response of the medium
resulting in absorption of the front part of the incident pulse larger
than that of its rear. The second model, referred to as the coherent-population-oscillations
(CPO) model, considers light beams whose intensity is slightly pulse
modulated and attributes the time delay of the transmitted pulse to
a reduction of the group velocity. We point out some inconsistencies
in the CPO model and show that the two models lie in reality on the
same hypotheses, the equations derived in the duly rectified CPO model
being local expressions of the integral equations obtained in the
incoherent bleaching model. When intense pulses without background
are used, the CPO model, based on linearized equations, breaks down.
The incoherent bleaching model then predicts that the transmitted
light should vanish when the intensity of the incident light is strictly
zero. This point is confirmed by the experiments that we have performed
on ruby with square-wave incident pulses and we show that the whole
shape of the observed pulses agrees with that derived analytically
by means of the incoherent bleaching model. We also determine in this
model the corresponding evolution of the fluorescence light, which
seems to have been evidenced in other experiments. 
\end{abstract}

\pacs{42.25.Bs, 42.25.Lc, 42.50.M}
\maketitle

\section{Introduction\label{sec:Introduction}}

The group velocity is a basic concept in the study of the propagation
of \emph{coherent} light pulses of slowly varying amplitude in a \emph{linear},
time independent, dispersive medium. It is invariably introduced in
every review on slow and fast light (see, e.g., \cite{bo02,bo09}).
Expressing that the phase of the optical field is stationary at the
frequency $\omega_{c}$ of the pulse carrier, it is given without
any ambiguity by the relation : 
\begin{equation}
v_{g}(\omega_{c})=\frac{c}{n(\omega_{c})+\omega_{c}dn/d\omega_{c}}\label{eq:equation0}
\end{equation}
where $c$ is the light velocity in vacuum, $n(\omega)$ designates
the refractive index at the optical frequency $\omega$ and $dn/d\omega_{c}$
is a short-hand notation of its derivative for $\omega=\omega_{c}$.
Slow light is obtained when $dn/d\omega_{c}$ takes large positive
values (steep dispersion), the group velocity being then much smaller
than the corresponding phase velocity $c/n(\omega_{c})$. As a consequence
of the Kramers-Kronig relations, this occurs in particular when the
carrier frequency $\omega_{c}$ of the light pulses is close to the
frequency $\omega_{0}$ of a well-marked maximum of the medium transmission
(minimum of absorption). When $\omega_{c}=\omega_{0}$, the group
velocity is minimal and a simple application of the moment theorem
shows that the center of gravity of the pulse envelope then exactly
propagates at the corresponding group velocity whatever the pulse
shape is \cite{ma03}. It should be mentioned that, due to the unavoidable
distortion of a pulse propagating in a dispersive absorptive medium,
the location $\tau_{max}$ of its maximum generally differs from that
$\tau_{g}$ of its center of gravity. It has been however demonstrated
that, under certain general conditions, the transmitted pulse becomes
nearly Gaussian for large enough propagation distances, with obviously
$\tau_{max}\approx\tau_{g}$ \cite{ma06}.

An important issue for eventual applications of slow light is the
fractional delay $\tau_{max}$/$\tau_{in}$ , where $\tau_{in}$ is
the full width at half-maximum (FWHM) of the intensity profile of
the incident pulse. Large fractional delays (up to 25 with moderate
pulse attenuation) have been evidenced when the required peak in the
medium transmission is associated with the minimum of absorption occurring
half-way between two absorption lines \cite{ta03,ca06,ca07}. Most
often, the transmission peak is obtained by applying an extra coherent
wave interacting nonlinearly with the medium, exploiting for instance
electromagnetically induced transparency (EIT) in atomic vapors \cite{ka95}
or Brillouin-induced gain in optical fibers \cite{so05}. Slow light
becomes a fashionable topic with the demonstration in an EIT experiment
of a group velocity as slow as $17\,\mathrm{m/s}$ in an ultra-cold
atomic gas \cite{ha99}. Subsequently the experiments showing delays
in the transmission of light pulses were systematically analyzed in
terms of slow group velocity. Such an analysis, indisputable for the
EIT experiments, is, however, questionable in the cases where the group
velocity as given by Eq.(1) is not well defined \cite{al06}.

We specifically examine here the case of the transmission
of light pulses through saturable absorbers. As far back as 1965,
Gires and Combaud \cite{gi65} showed that the stationary transmission
of organic dyes is fairly well reproduced by assimilating the medium
to a resonant two-level medium and using the rate equations approximation.
By this means they obtained two equations coupling the population
difference and the beam intensity. Solving these equations in the
time-dependent case, Selden theoretically studied the transmission
of light pulses, demonstrating narrowing, skewing, and time delay of
the transmitted pulse \cite{se67,se70} in agreement with the experimental
observations \cite{kh67,he68,se69}. From a qualitative viewpoint,
the delay is thus interpreted in terms of \emph{pulse reshaping},
the leading edge of the incident pulse being more attenuated than
its trailing edge. Insofar as only the intensity is involved in the
process, the phenomenon is currently referred to as \emph{incoherent
bleaching} to distinguish it from the pulse delay and reshaping that
occur in purely coherent cases. Smith and Allen showed that delays
sometimes attributed to self-induced transparency (a \emph{coherent
}phenomenon) are in fact the result of \emph{incoherent} bleaching
and are well reproduced by Selden's theory \cite{sm73}. For a
complete analysis of this point, see \cite{al87}. Always by using
the incoherent bleaching model, Selden also examined the transmission
of a continuous wave (cw) whose intensity is slightly modulated by
a sine wave \cite{se71}. He showed that the intensity modulation
index increases with the propagation distance whereas its phase is
time delayed. These phenomena were observed on a ruby crystal at room
temperature by Hillmann \emph{et al.} \cite{hi83} and by Bigelow
\emph{et al}. \cite{bi03}. The latter, not taking into account the results of the
incoherent bleaching model and considering the time delay of the modulation
phase to be a group delay, claimed to have discovered ``a new method
that produces slow propagation of light\textquotedblright{} \cite{bi03}.
Their theoretical analysis was based on an extrapolation of the results
obtained when two separated \emph{coherent} waves originate \emph{coherent}
oscillations of the populations at their frequency difference. Abundantly
cited, Ref. \cite{bi03} paved the way to numerous articles invoking
slow light based on coherent population oscillations (CPO). The systems
under consideration are very various, comprising in particular doped
crystals \cite{bi04,ba05,bi06}, semi-conductor devices \cite{ku04,va05,mo05,su06,ku07},
doped optical fibers \cite{sc06,ar10,be10,ja15} and doped glass microspheres
\cite{hu16}. However, as shown in \cite{al06,za06,ma08,se09}, the
effects reported in most of these articles do not involve coherence
in the optical sense and can be explained in the frame of the incoherent
bleaching model. The controversy on this matter restarted more recently
with the publication of an article reporting experiments performed
with a spinning ruby window \cite{wi13}. In this article Wisniewski-Barker
\emph{et al}. claim that their results are ``incompatible with slow-light
models based on simple pulse-reshaping arising from optical bleaching. \textquotedblright This statement was contested by Kozlov \emph{et al.} \cite{ko14}
who performed an experiment validating the incoherent bleaching model
but, surprisingly enough, Wisniewski-Barker \emph{et al.} obtained
the opposite result by using practically the same experimental setup
\cite{wi14}. The controversy between the two models thus remains
open. We attempt in the following to solve this issue. In Sec. \ref{sec:Weakmodulation},
we revisit the case of weak modulation depths by extending the results
given in \cite{se71,ma08} and showing that, after correction, the
CPO equations are simply local expressions of the integral equations
obtained in the incoherent bleaching model. In Sec. \ref{sec:IntensePulse},
we examine the validity of the bleaching model in the case of saturating
pulses without background, we report experiments confirming the positive
result of Kozlov \emph{et al}. and we give a possible explanation
of the different result reported in \cite{wi14}. We finally conclude
in Sec. \ref{sec:Conclusion} by summarizing and discussing our main
results.

\section{Case of weak modulation depths\label{sec:Weakmodulation}}

The CPO model traces back to the paper of Schwarz and Tan who studied
how the absorption of a coherent probe wave by a saturable absorber
is modified when the medium is submitted to a coherent saturating
wave \cite{sc67}. They showed that the probe absorption spectrum
then displays a dip (hole) centered at the pump frequency and of width
$\approx1/T_{1}$, where $T_{1}$ is the population relaxation time.
This dip is considered in \cite{hi83} as resulting from the population
oscillations created in the medium by the beating of the pump and
probe waves. When the directions of propagation of the two waves are
different, it is possible to determine without ambiguity the absorption
coefficient, the refractive index, and the group velocity of the probe
wave. An experiment corresponding to this scheme has been performed
by Ku \emph{et al.}, the slow group velocity being inferred from a
measurement of the phase of the probe field \cite{ku04}. A related
experiment involving counter-propagative waves is reported in \cite{ja15}.

The problems arise when the previous results are extended to the study
of a single cw whose intensity is slightly modulated by a sine wave
of low frequency. Denoting $\Omega$ the modulation frequency, the
modulated wave can be considered as the superimposition of three co-propagative
cws, a saturating wave of frequency $\omega_{s}$ and two sidebands
of frequency $\omega_{s}\pm\Omega$, acting as probes \cite{re1}.
Bigelow \emph{et al.} consider in \cite{bi03} that the two probes
act independently but, as soundly noted by Sargent, ``although neither
probe frequency could influence the other on its own, they succeed
in doing so with the help of the saturating wave\textquotedblright{}
\cite{sa78}. This indicates that Eq. (9) in \cite{bi03} is not correct. This error
has been pointed out for the first time by Mørk\emph{ et al.} \cite{mo05}
who indicated that a correct application of the four-wave mixing theory
leads to multiply the phase lag of the modulation and thus the corresponding
time delay by a factor two.

Another point raised by Zapasskii and Kozlov \cite{za06} is that
the CPO model implicitly assumes that the saturating wave has a spectral
width much smaller than that of the hole induced in the absorption
spectrum ($\approx1/T_{1}$). This condition is far from being met
in most of the experiments. There is then no dip in the \emph{optical}
absorption spectrum and thus no associated slow light in the usual
sense. In response to this objection, Piredda and Boyd \cite{pi07},
while continuing to invoke CPO, developed a model based on two equations
coupling wave intensity and ground state population identical to those
coupling wave intensity and population difference in the incoherent
bleaching model \cite{gi65,se67,ma08}. They can be written in the
simplified form

\begin{equation}
T_{1}\frac{\partial N}{\partial t}=-N\left(1+I\,\right)+1,\label{eq:equation1}
\end{equation}
\begin{equation}
\frac{\partial I}{\partial z}=-\alpha IN,\label{eq:equation2}
\end{equation}
where $T_{1}$ is the ground-state recovery time, $N$ is the ground-state population normalized to its value at equilibrium, $t$ is the time retarded by the propagation time in the host medium (negligible
compared to the delays considered in the following), $I$ is the beam
intensity normalized to the saturation intensity, $z$ is the abscissa
along the direction of propagation, and $\alpha$ is the unsaturated
absorption coefficient. Equations (\ref{eq:equation1}) and (\ref{eq:equation2})
implicitly assume that the ground state has no resolved structure.
This condition is met in saturable absorbers which are dense media
working at room temperature. It prevents oscillations at the frequency
of a transition between two separated sublevels as evidenced in an
alkali-metal vapor in buffer gas \cite{go02,re3}. In the case of a cw,
Eqs (\ref{eq:equation1}) and (\ref{eq:equation2}) are reduced to 
\begin{equation}
\bar{N}\left(z\right)=\frac{1}{1+\bar{I}(z)},\label{eq:equation3}
\end{equation}
\begin{equation}
\frac{\partial\bar{I}\left(z\right)}{\partial z}=-\frac{\alpha\bar{I}(z)}{1+\bar{I}(z)},\label{eq:equation4}
\end{equation}
\begin{equation}
\bar{I}(z)+\ln\bar{I}(z)=\bar{I}(0)+\ln\bar{I}(0)-\alpha z,\label{eq:equation5}
\end{equation}
where, as in the following, upper bars refer to time-independent quantities.
When the cw is slightly modulated, $I(z,t$)=$\bar{I}(z)+\Delta I(z,t)$
with $\Delta I(z,t)\ll\bar{I}(z)$ and $N(z,t$)=$\bar{N}(z)-\Delta N(z,t)$
with $\Delta N(z,t)\ll\bar{N}(z)$ . Making a calculation at the first
order in $\Delta I$ and $\Delta N$, taking into account Eqs.(\ref{eq:equation3}) and (\ref{eq:equation5}) and passing in the Fourier space \cite{pa87},
we obtain the transfer functions $H(z,\Omega)$ relating the Fourier
transform $\Delta I(z,\Omega)$ of $\Delta I(z,t)$ to that $\Delta I(0,\Omega)$
of $\Delta I(0,t)$ and $H_{\Delta N}(z,\Omega)$ relating the Fourier
transform $\Delta N(z,\Omega)$ of $\Delta N(z,t)$ to $\Delta I(z,\Omega)$.
They read 
\begin{multline}
H(z,\Omega)=\left(\frac{\bar{I}(z)}{\bar{I}(0)}\right)\left(\frac{1+\bar{I}(0)+i\Omega T_{1}}{1+\bar{I}(z)+i\Omega T_{1}}\right)\\
=\left(\frac{\bar{I}(z)}{\bar{I}(0)}\right)\left(1+\frac{\bar{I}(0)-\bar{I}(z)}{1+\bar{I}(z)+i\Omega T_{1}}\right).\label{eq:EquationSix}
\end{multline}
\begin{equation}
H_{\Delta N}(z,\Omega)=\frac{1}{\left[1+\bar{I}(z)\right]\left[1+\bar{I}(z)+i\Omega T_{1}\right]}.\label{eq:equation7}
\end{equation}
The transfer function $H(z,\Omega)$ has a single pole and a single
zero, both purely imaginary with a positive imaginary part. This implies
that its inverse Fourier transform, that is the system impulse response,
is real and that the system is causal with minimum phase shift \cite{pa87}.
The phase $\Phi(z,\Omega)$ of $H(z,\Omega)$ then obeys to the relation
\begin{equation}
\Phi\left(z,\Omega\right)=-\mathcal{H}\left\{ \left|\ln\left(H\left(z,\Omega\right)\right)\right|\right\},\label{eq:equation8}
\end{equation}
where $\mathcal{H}$ designates the Hilbert transform. Equation (\ref{eq:equation8})
may be considered as a generalized Kramers-Kronig relation. Similar
properties hold for $H_{\Delta N}(z,\Omega)$. We incidentally remark
that $H(z,\Omega)$ is the transfer function of a simple electric
network involving an RC circuit, two voltage dividers, and a voltage
adder whereas an RC circuit and one voltage divider suffice to reproduce
$H_{\Delta N}(z,\Omega)$.

To relate Eq. (\ref{eq:EquationSix}) to the results given by Piredda
and Boyd in \cite{pi07}, we consider the transfer function $H(dz,\Omega)=H(z+dz,\Omega)/H(z,\Omega)$
of the infinitely thin slice comprised between $z$ and $z+dz$ in
the medium. Taking into account Eq.(\ref{eq:equation4}), we obtain:

\begin{equation}
H(dz,\Omega)=\exp\left[-\alpha_{mod}\left(z,\Omega\right)dz-i\varphi_{mod}\left(z,\Omega\right)dz\right],\label{eq:equation9}
\end{equation}
with 
\begin{equation}
\alpha_{mod}\left(z,\Omega\right)=\frac{\alpha}{1+\bar{I}(z)}\left\{ 1-\frac{\bar{I}(z)\left[1+\bar{I}(z)\right]}{\left[1+\bar{I}(z)\right]^{2}+\left(\Omega T_{1}\right)^{2}}\right\}, \label{eq:equationDix}
\end{equation}
\begin{equation}
\varphi_{mod}\left(z,\Omega\right)=\frac{\alpha\bar{I}(z)}{1+\bar{I}(z)}\left\{ \frac{\Omega T_{1}}{\left[1+\bar{I}(z)\right]^{2}+\left(\Omega T_{1}\right)^{2}}\right\}. \label{eq:equationOnze}
\end{equation}
In relation with Eq. (\ref{eq:equation8}), we note that $\varphi_{mod}\left(z,\Omega\right)=\mathcal{H}\left[\alpha_{mod}\left(z,\Omega\right)\right]$.
When the modulation is reduced to a sine wave of frequency $\Omega$
(as considered in most experiments), $\alpha_{mod}\left(z,\Omega\right)$
is the attenuation coefficient of the modulation and $\varphi_{mod}\left(z,\Omega\right)$
is the associated phase lag per length unit. Equations.(\ref{eq:equationDix}) and
(\ref{eq:equationOnze}) are strictly equivalent to those obtained
in \cite{pi07} by invoking the CPO model and this shows that \emph{the
CPO model does not bring new results with regard to the incoherent
bleaching model}. As expected, the phase lag is two times that given
in \cite{bi03,re2}. A phase velocity $v_{\varphi mod}(z,\Omega)=\Omega/\varphi_{mod}\left(z,\Omega\right)$
can be associated with this phase lag. Depending on $\Omega$,\emph{
this velocity should not be confused with a group velocity} as it
is made in \cite{bi03,pi07}. When the modulation consists in a pulse,
the modulation group delay $d\tau_{gmod}$ through the slice $(z,z+dz)$
is derived from $H(dz,\Omega)$ by the moment theorem. The
slice being assumed to be infinitely thin $H(dz,0)\approx1$ and $d\tau_{gmod}$
can be identified to the coefficient of the first degree term in the
expansion of $H(dz,\Omega)$ in power series of $\left(-i\Omega\right)$.
We get
\begin{equation}
d\tau_{gmod}=\frac{\alpha\bar{I}(z)T_{1}dz}{\left[1+\bar{I}(z)\right]^{3}},\label{eq:equationDouze}
\end{equation}
and the corresponding (local) group velocity $v_{gmod}\left(z\right)=dz/d\tau_{gmod}$
reads: 
\begin{equation}
v_{gmod}\left(z\right)=\frac{\left[1+\bar{I}(z)\right]^{3}}{\alpha\bar{I}(z)T_{1}},\label{eq:equationTreize}
\end{equation}
Note that this group velocity is related to the intensity modulation
transmission and should be distinguished from the group velocity as
defined by Eq.(\ref{eq:equation0}) for pulses of coherent light.

\begin{figure}[h]
\begin{centering}
\includegraphics[width=0.95\columnwidth]{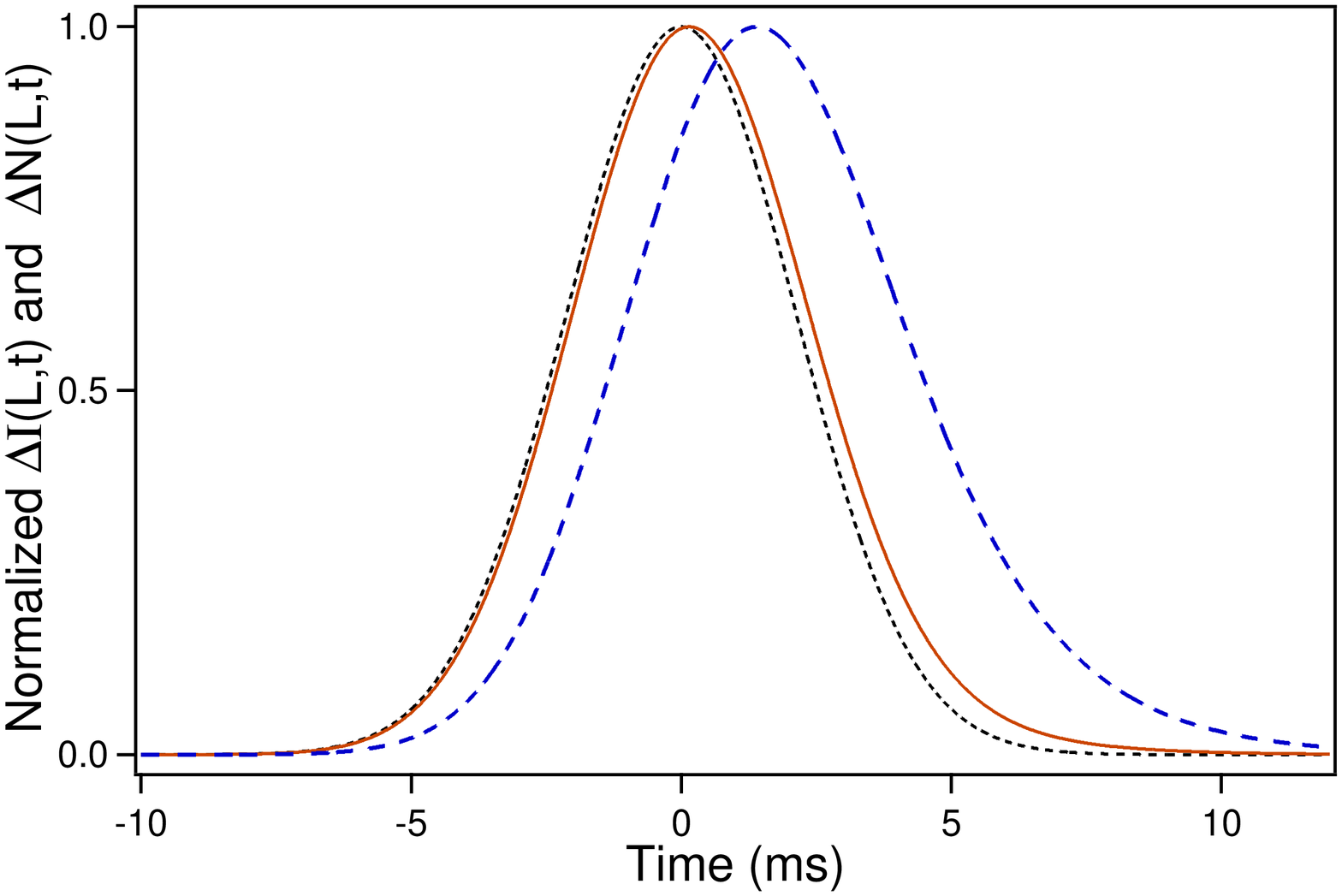} 
\par\end{centering}
\caption{Normalized transmitted pulse $\Delta I\left(L,t\right)$ (solid line)
and population variation $\Delta N\left(L,t\right)$ (dashed line).
Parameters: $\bar{I}\left(0\right)=0.23$, $T_{1}=1.6\,\mathrm{ms}$
, $\alpha=1.17\,\mathrm{cm^{-1}}$ and $L=4.25\,\mathrm{cm}$ , leading
to $\alpha L\approx5$, $\bar{I}(L)\approx2\times10^{-3}$, $T_{L}\approx T_{1}$,
$\Delta\tau_{g}(L)\approx T_{1}$ and $\tau_{gmod}(L)\approx0.3\,\mathrm{ms}$.
The normalized incident pulse $\Delta I\left(0,t\right)$ (dotted
line) of FWHM duration $\tau_{in}=5\,\mathrm{ms}$ is given for reference.
Note the large delay of $\Delta N\left(L,t\right)$ compared to that
of $\Delta I\left(L,t\right)$.\label{fig:Figure1}}
\end{figure}

The attenuation of the cw and of the modulation being intrinsically
coupled, the widespread approximation consisting in neglecting the
former to study the latter is not justified. The medium being
assumed to occupy the space $0\leq z\leq L$, the use of the integral
expressions of Eq.(\ref{eq:EquationSix},\ref{eq:equation7}) has
the advantage to give directly $\Delta I(L,\Omega)$ and $\Delta N(L,\Omega)$
without requiring integration in $z$. The corresponding expressions
of $\Delta I(L,t)$ and $\Delta N(L,t)$ read 
\begin{equation}
\varDelta I\left(L,t\right)=\mathcal{F}^{-1}\left[H(L,\Omega)\varDelta I\left(0,\Omega\right)\right],\label{eq:equationQuatorze}
\end{equation}
\begin{equation}
\varDelta N\left(L,t\right)=\mathcal{F}^{-1}\left[H_{\varDelta N}\left(L,\Omega\right)H(L,\Omega)\varDelta I\left(0,\Omega\right)\right],\label{eq:equationQuinze}
\end{equation}
where $\mathcal{F}^{-1}$ designates inverse Fourier transforms. General
characteristics of $\Delta I(L,t)$ and $\Delta N(L,t)$ can be derived
by exploiting the remarkable properties of the cumulants \cite{do03,ma06}.
The cumulants $\kappa_{n}$ of the Fourier transform $G(\Omega)$
of a real function $g(t)$ are given by the expansion 
\begin{equation}
G\left(\Omega\right)=G\left(0\right)\exp\left(\sum_{n=1}^{\infty}\frac{\kappa_{n}}{n!}\left(-i\Omega\right)^{n}\right)\label{eq:equationSeize}
\end{equation}
where the cumulants $\kappa_{1}$, $\kappa_{2}$ and $\kappa_{3}$
can be shown to be respectively equal to the mean time of $g(t)$,
its variance $\sigma^{2}$, and its third centered moment $\mu_{3}$.
For $G\left(\Omega\right)=H(L,\Omega)$, we get $\kappa_{1}=T_{L}-T_{0}$,
$\kappa_{2}=T_{L}^{2}-T_{0}^{2}$ and $\kappa_{3}=2(T_{L}^{3}-T_{0}^{3})$
where $T_{z}$ is a short hand notation of $T_{1}/[1+\bar{I}(z)]$.
When the modulation is pulsed, the additivity property of the cumulants
enables us to identify $\kappa_{1}$ to the time delay of the pulse
center-of-gravity (modulation group delay), $\kappa_{2}$ to the increase
of the pulse variance, and $\kappa_{3}$ to that of $\mu_{3}$. The
group delay for the whole medium thus reads 
\begin{equation}
\tau_{gmod}\left(L\right)=\frac{T_{1}}{1+\bar{I}\left(L\right)}-\frac{T_{1}}{1+\bar{I}\left(0\right)}.\label{eq:equationDixsept}
\end{equation}
We incidentally remark that this result can be retrieved by an integration
combining Eqs.(\ref{eq:equationDouze}) and (\ref{eq:equation4}).
An important point is that the group delay cannot exceed $T_{1}$
as large as the medium thickness may be. This limit is attained when
$\bar{I}(0)\gg1$ (strongly saturating incident cw) and $\bar{I}(L)\ll1$
(medium nearly opaque). The variance and the third moment of the transmitted
pulse are then also the largest. If the incident pulse is symmetric,
the large positive value of $\kappa_{3}$ entails that the output
pulse will be strongly skew with a rise much steeper than its fall
and, consequently, a time delay $\tau_{max}$ of its maximum much
shorter than the group delay $\tau_{gmod}$. See, for illustration,
Fig.2(c) in \cite{ma08}. In the more realistic case where $\bar{I}(0)<1$
[always with $\bar{I}(L)\ll1$], the pulse distortion will be moderate
but the group delay will be short compared to $T_{1}$ (while keeping
significantly longer than $\tau_{max}$). When the incident pulse
is a Gaussian of the form $\varDelta I\left(0,t\right)\propto\exp\left(-t^{2}/\tau_{p}^{2}\right)$,
the transmitted pulse derived from Eq.(\ref{eq:equationQuatorze})
reads 
\begin{multline}
\varDelta I\left(L,t\right)\propto\varDelta I\left(0,t\right)\\
+\frac{\bar{I}\left(0\right)\tau_{p}\sqrt{\pi}}{2T_{1}}\left[1+erf\left(\frac{t}{\tau_{p}}-\frac{\tau_{p}}{2T_{1}}\right)\right]\exp\left(\frac{\tau_{p}^{2}}{4T_{1}^{2}}-\frac{t}{T_{1}}\right),\label{eq:equationDixhuit}
\end{multline}
where erf designates the error function. Figure \ref{fig:Figure1}
shows the result obtained in the conditions of the experiment on ruby
reported in \cite{bi06}. The parameters are $\bar{I}\left(0\right)=0.23$,
$T_{1}=1.6\,\mathrm{ms}$ , $\alpha=1.17$ $\mathrm{cm^{-1}}$, and $L=4.25\,\mathrm{cm}$,
from which we deduce $\alpha L\approx5$, $\bar{I}(L)\approx2\times10^{-3}$,
$T_{L}\approx T_{1}$ and $\tau_{gmod}(L)\approx0.3\,\mathrm{ms}$
(actually much shorter than $T_{1}$). 

The incident pulse (dotted line) has a FWHM duration $\tau_{in}=2\tau_{p}\sqrt{\ln2}=5\,\mathrm{ms}$.
As predicted, $\tau_{max}$ is significantly shorter than $\tau_{gmod}$
($\tau_{max}\approx\tau_{gmod}/2$ ) . We also compare in Fig. \ref{fig:Figure1}
$\Delta N(z,L)$ to $\Delta I(z,L)$. It is easily deduced from Eq.(\ref{eq:equation7})
that the group delay of $\Delta N(z,L)$ exceeds that of $\Delta I(z,L)$
by a quantity 
\begin{equation}
\varDelta\tau_{g\Delta N}=\frac{T_{1}}{1+\bar{I}\left(L\right)}\approx T_{1}\label{eq:equationDixneuf}
\end{equation}
An important point is that this extra delay of the population evolution
is much longer than the delay of the signal $\Delta I(z,L)$. 
\begin{figure}[h]
\begin{centering}
\includegraphics[width=0.95\columnwidth]{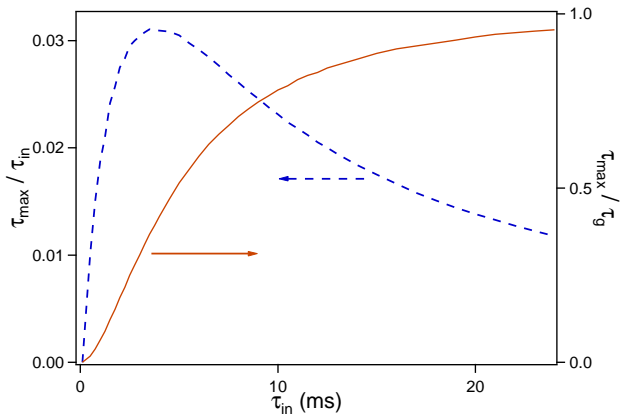} 
\par\end{centering}
\caption{Fractional delay $\tau_{max}/\tau_{in}$ of the transmitted pulse
(dashed line, left scale) and ratio $\tau_{max}/\tau_{gmod}$ (solid
line, right scale) as functions of the duration $\tau_{in}$ of the
incident pulse. Other parameters as in Fig. \ref{fig:Figure1}. \label{fig:Figure2}}
\end{figure}

Coming back to the latter, Fig. \ref{fig:Figure2} shows how the fractional
delay $\tau_{max}/\tau_{in}$ of the transmitted pulse and the ratio
$\tau_{max}/\tau_{gmod}$ depend on the duration $\tau_{in}$ of the
incident pulse. It appears that $\tau_{max}$ approaches its asymptotic
value $\tau_{gmod}$ for values of $\tau_{in}$ at which the fractional
delay tends to 0 and, conversely, that the latter attains its maximum
for a pulse duration $\tau_{in}$ such that is only about $\tau_{gmod}/3$
.

\section{Case of intense pulses without background\label{sec:IntensePulse}}

Even duly rectified, the CPO model, based on linearized equations,
does not apply to the case of saturating pulses without background.
In order to extend its range of application to such situations, its
defenders invoke a mechanism of ``self-pumping\textquotedblright ,
one part of the pulse acting as a pump whereas the remaining part
acts as the probe \cite{bi03,wi14}. Without any quantitative support,
this claim seems purely incantatory. It is even qualitatively incompatible
with the fact that smooth symmetric pulses are broadened and gain
a positive skewness (rise steeper than the fall) when they are superimposed
on a large background (range of validity of the CPO model) whereas
saturating pulses without background are narrowed and gain a negative
skewness (fall steeper than the rise) \cite{se67,kh67,ma08}. Reporting
experiments performed with a spinning ruby window, Wisniewski-Barker
\emph{et al.} \cite{wi13} recently proclaimed the failure of the
incoherent bleaching model for this reference material. According
to the demonstration made in the previous section of the equivalence
of the two models, all the results obtained by means of the CPO model
in the weak modulation limit would then require revision. Fortunately
enough, it is nothing of the sort. The argument given in \cite{wi13}
against the incoherent bleaching model is that, in this model, the
transmitted light intensity should vanish when the incident light
intensity is null, a condition that would be not verified in the experiments.
In a comment, Kozlov \emph{et al. }\cite{ko14} contested the achievement
of strictly null incident intensity in these experiments and proposed
a more drastic test where the incident beam is switched on-off by
a mechanical chopper (100\% square wave modulation). Performing this
experiment, they obtained results validating the incoherent bleaching
model, without the slightest transmitted intensity after the switching
off of the incident beam. This however did not close the debate. Indeed,
accepting the challenge proposed by Kozlov \emph{et al.}, Wisniewski-Barker\emph{
et al.} \cite{wi14} performed a nearly identical experiment. They
obtained opposite results, evidencing in particular an exponential
decrease of the transmitted intensity after the switching off of the
incident beam. Note that they did not attempt to explain the discrepancy
between their results and those of Kozlov\emph{ et al.} To solve this
issue, we describe in the following an experiment that validates the
incoherent bleaching model and theoretically determine in this model
the time dependence of $I\left(L,t\right)$ and of $N\left(L,t\right)$
, providing a possible explanation of the experimental results obtained
in \cite{wi14}. 
\begin{figure}[h]
\begin{centering}
\includegraphics[width=0.95\columnwidth]{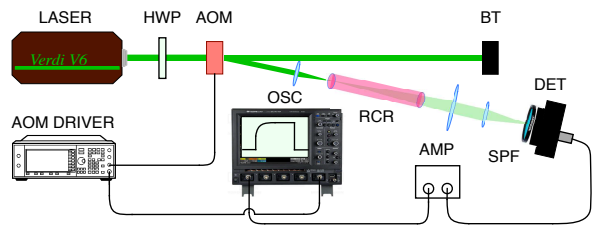} 
\par\end{centering}
\caption{Experimental set-up. LASER : single longitudinal mode Nd: YVO4 laser
(Verdi V6, Coherent) delivering a cw at 532 nm; HWP, half-wave plate;
AOM, acousto-optic modulator (AA Opto-Electronic, MTS110-A-VIS);
BT, beam trap absorbing the non-diffracted light; RCR, ruby crystal
rod; SPF, dichroic shortpass filter (Semrock: BSP-633R-25) eliminating
the ruby fluorescence light at 694 nm; DET, high-speed silicon photodetector
(Thorlabs DET 210, DC-350MHz); AMP, low-band (DC-200 kHz) amplifier
(Hamamatsu C7319); OSC, digital oscilloscope (Lecroy Waverunner 104
MXi); AOM DRIVER, function generator driving the acousto-optic modulator
and triggering the oscilloscope.\label{fig:Setup}}
\end{figure}

Figure \ref{fig:Setup} shows our experimental setup. It is
very similar to those used in \cite{ko14,wi14}. We use a single longitudinal
mode Nd:YVO$_{4}$ laser (Verdi V6, Coherent) operating at $532\,\mathrm{nm}$
as a cw source of controllable power. The laser beam is collimated
and sent on an acousto-optic modulator (AA Opto-Electronic, MTS110-A-VIS).
The light polarization at the input of the modulator is adjusted by
a half-wave plate. A sine wave, at a 110 MHz acoustic frequency, is
applied to the modulator crystal. Its amplitude is driven by a function
generator which enables us to switch the power diffracted in the first
order from 0 to 80\% of the power delivered by the laser. The switching
times are about $1\,\mathrm{\mu s}$. A lens of 50 mm focal length
focuses the diffracted beam slightly behind the front face of a 120
mm long standard laser ruby crystal rod. The transmitted beam is focused
onto a high-speed silicon photodetector (Thorlabs DET 210, DC-350
MHz). The detector is preceded by a dichroic shortpass filter (Semrock:
BSP-633R-25) of optical density exceeding 7 at 694 nm, which eliminates
the fluorescence light emitted by the ruby crystal. The signal delivered
by the detector is amplified by a low-band (DC-200 kHz) amplifier
(Hamamatsu C7319) and averaged by a digital oscilloscope (Lecroy Waverunner
104 MXi). We have carefully verified that the power diffracted in
the first order is null when the modulator is in the off position
(amplitude of the acoustic wave equal to zero) and that the corresponding
signal delivered by the low-band amplifier vanishes. Numerous experiments
have been performed for various laser powers and for several positions
of the minimum beam waist inside the ruby crystal. In all these experiments,
we never observe any detected signal when the power is switched off.
\begin{figure}
\centering{} \includegraphics[width=0.95\columnwidth]{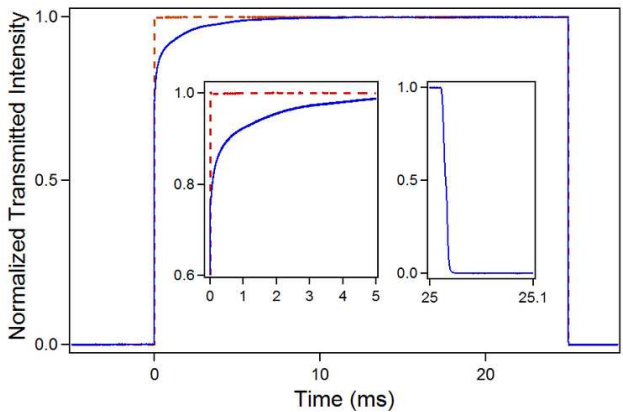}
\caption{Example of transmitted signal observed in our experiments for a power
sent on the ruby crystal of about $1.6\,\mathrm{W}$ and a pulse duration
$\tau_{in}=25\,\mathrm{ms}$ (solid line). The dashed line is the
profile of the incident pulse obtained by replacing the ruby crystal
by a suitable neutral density filter. The insets are enlargements
of the pulse rise and fall. \label{fig:Figure3}}
\end{figure}

Figure \ref{fig:Figure3} gives an example of detected signal obtained
in such conditions. The power applied on the front face of the ruby
crystal is modulated from 0 to 1.6 W by a 20-Hz square wave. The leading
edge of the signal is characterized by an almost instantaneous jump
which brings the signal to a value equal to 75\% of its maximum value.
This rapid variation is followed by a nearly exponential slow increase
with a time-constant of about $1.6\,\mathrm{ms}$. In accordance with
the incoherent bleaching model, on its trailing edge the signal quickly
returns to zero with a 90\%-10\% switching time $<6\,\mathrm{\mu s}$
mainly introduced by the amplifier. Note that, for completeness, we
have also performed experiments where the acousto-optic modulator
was replaced by a mechanical chopper as in \cite{ko14,wi14}. The
results obtained with both modulators are identical. Finally, although
this parameter is not critical, we have measured the unsaturated optical
thickness of the ruby crystal by collimating the beam inside the rod.
The linear evolution of the transmitted power versus the input one
leads to a mean value of $\alpha L$ close to 9.

To analyze the transmitted signals obtained in our experiments as
those reported in \cite{ko14,wi14}, we come back to Eqs.(\ref{eq:equation1},
\ref{eq:equation2}) which may be considered as the basis of the incoherent
bleaching model. They first show that $I\left(L,t\right)=0$ when
$I\left(0,t\right)=0$ . Analytical results valid at every time can
be obtained when the medium thickness is such that $I\left(L,t\right)\ll I\left(0,t\right)$
, a condition met in the experiments. As in \cite{se67,ma08}, we
introduce the function 
\begin{equation}
Z(L,t)=\ln\left[I(L,t)\right]-\ln\left[I(0,t)\right]+\alpha L.\label{eq:equationVingt}
\end{equation}
From Eqs.(\ref{eq:equation1},\ref{eq:equation2}), we deduce 
\begin{equation}
T_{1}\frac{dZ}{dt}+Z=I(0,t)-I(L,t)\approx I(0,t),\label{eq:equationVingtetUn}
\end{equation}
and 
\begin{equation}
I(L,t)\approx I(0,t)\exp\left[Z(L,t)-\alpha L\right].\label{eq:equationVingtdeux}
\end{equation}
In the case of a square incident pulse of the form $I\left(0,t\right)=I_{0}\left[u_{H}\left(t\right)-u_{H}\left(t-\tau_{in}\right)\right]$
where $u_{H}\left(t\right)$ designates the Heaviside unit step function
and $\tau_{in}$ is the pulse duration, we get 
\begin{equation}
I(L,t)\approx I(0,t)\,e^{-\alpha L}\exp\left[I_{0}\left(1-e^{-t/T_{1}}\right)\right].\label{eq:equationVingttrois}
\end{equation}
The transmitted pulse displays an initial discontinuity equal to $I(0,t)\,e^{-\alpha L}$
at $t=0$ before rising as an exponential of exponential and falling
to 0 at $t=\tau_{in}$ . If $\tau_{in}\gg T_{1}$, it attains the
asymptotic limit $I_{0}\exp\left(I_{0}-\alpha L\right)$ before falling.
Note also that the rise is reduced to a simple exponential if $I_{0}\ll1$.
We have then 
\begin{equation}
I(L,t)\approx I(0,t)\,e^{-\alpha L}\left[1+I_{0}\left(1-e^{-t/T_{1}}\right)\right].\label{eq:equaionVingtQuatre}
\end{equation}
\begin{figure}
\begin{centering}
\includegraphics[width=0.95\columnwidth]{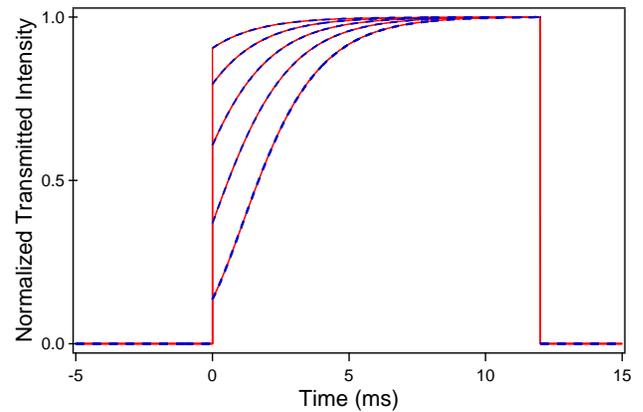} 
\par\end{centering}
\caption{Theoretical transmitted intensity as a function of time for an incident
square pulse. Parameters: $T_{1}=1.6\,\mathrm{\mu s}$, $\alpha L=9$,
$\tau_{in}=12\,\mathrm{ms}$ and, from top to bottom, $I_{0}=0.1,\,0.23,\,0.5,\,1$
and $2$. The solid and dashed lines are respectively the exact numerical
solution and the approximate analytic solution given by Eq. (\ref{eq:equationVingttrois}).\label{fig:Figure4}}
\end{figure}

Figure \ref{fig:Figure4} shows the results obtained for $T_{1}=1.6\,\mathrm{\mu s}$,
$\alpha L=9$, $\tau_{in}=12\,\mathrm{ms}$ and, from top to bottom,
$I_{0}=0.1,\,0.23,\,0.5,\,1$ and $2$. Note that the analytical solution
given by Eq. (\ref{eq:equationVingttrois}) perfectly fits the exact
numerical solution as long as $I_{0}\ll\alpha L$ and that the shape
of the transmitted pulse then does not depend on $\alpha L$. The
values $I_{0}=0.23$ and $T_{1}=1.6\,\mathrm{\mu s}$ approximately
correspond to the experiment of Fig. \ref{fig:Figure3} and Eq. (\ref{eq:equaionVingtQuatre})
satisfactorily holds in this case.

In order to determine the normalized population $N_{F}\approx1-N$
of the fluorescent metastable level (again in the limit $I_{0}\ll\alpha L$
), we come back to Eq. (\ref{eq:equationVingtetUn}) and replace $I\left(L,t\right)$
by its approximate form given by Eqs.(\ref{eq:equationVingttrois}).
After a tedious calculation, we get for the rise of $N_{F}\left(L,t\right)$
the rather complex expression 
\begin{equation}
N_{F}\left(L,t\right)=I_{0}e^{-\alpha L}f(t)\,e^{-t/T_{1}}u_{H}(t),\label{eq:equationVingtcinq}
\end{equation}
with 
\begin{multline}
f(t)=\exp\left[\frac{t}{T_{1}}+I_{0}\left(1-e^{-t/T_{1}}\right)\right]\\
+I_{0}e^{\,I_{0}}\left[E_{1}\left(I_{0}\right)-E_{1}\left(I_{0}e^{-t/T_{1}}\right)\right]-1,\label{eq:equationVingtsix}
\end{multline}
where $E_{1}(x)$ designates the exponential integral function \cite{ni10}.
Equations. (\ref{eq:equationVingtcinq}) and (\ref{eq:equationVingtsix}) hold
for arbitrary $I_{0}\ll\alpha L$. For $t>\tau_{in}$, it immediately
results from Eq. (\ref{eq:equation2}) that, without any restriction
on $I_{0}$ and $\alpha L$, $N_{F}\left(L,t\right)$ is reduced to
a decreasing exponential of time constant $T_{1}$, namely 
\begin{equation}
N_{F}\left(L,t\right)=N_{F}\left(L,\tau_{in}\right)\exp\left(-\frac{t-\tau_{in}}{T_{1}}\right)u_{H}(t-\tau_{in}).\label{eq:equationVingtSept}
\end{equation}
\begin{figure}
\begin{centering}
\includegraphics[width=0.95\columnwidth]{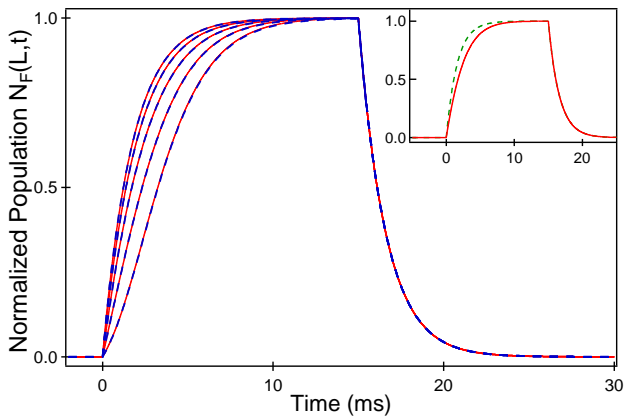} 
\par\end{centering}
\caption{Theoretical evolution of the fluorescent metastable level population.
Pulse duration $\tau_{in}=15\,\mathrm{ms}$. Other parameters as in
Fig. \ref{fig:Figure4}. The solid and dashed lines are respectively
the exact numerical solution and the analytic solution given by Eqs. (\ref{eq:equationVingtcinq})-(\ref{eq:equationVingtSept}).
Inset: Comparison for $I_{0}=0.5$ of the exact numerical solution
(solid line) to that given by Eqs. (\ref{eq:equationVingtSept}) and (\ref{eq:equationVingthuit})
analogous to those of a RC circuit (dashed line).\label{fig:Figure5}}
\end{figure}

Figure \ref{fig:Figure5} shows the evolution of the population of
the fluorescent metastable level obtained for the intensities already
considered in Fig.\ref{fig:Figure4}. We see that Eqs. (\ref{eq:equationVingtcinq})-(\ref{eq:equationVingtSept})
perfectly fit the exact numerical solutions. When $I_{0}\ll1$, Eq.(\ref{eq:equationVingtcinq})
becomes at the lowest order in $I_{0},$ 
\begin{equation}
N_{F}\left(L,t\right)=I_{0}e^{I_{0}-\alpha L}\left[1-e^{-t/T_{1}}\right]u_{H}(t)\label{eq:equationVingthuit}
\end{equation}
Equations.(\ref{eq:equationVingthuit}) and (\ref{eq:equationVingtSept})
are identical to those describing respectively the charge and the
discharge of a capacity $C$ through a resistance $R$ with $RC=T_{1}$.
As shown in the inset of Fig. \ref{fig:Figure5}, they provide a satisfactory
approximation of the exact result for $I_{0}$ as large as $0.5$.

Our experimental results on the transmitted pulse $I\left(L,t\right)$
obtained for an incident square pulse confirms those obtained by Kozlov
\emph{et al}. \cite{ko14} and are in good agreement with our calculations
based on the incoherent bleaching model. The key points are (i) $I\left(L,t\right)$
presents a discontinuity followed by an exponential-like rise when
the incident beam is switched on (ii) $I\left(L,t\right)$ immediately
falls to 0 when the incident beam is switched off. Quite different
results are reported by Wisniewski \emph{et al}. \cite{wi14}: There
is no initial discontinuity of $I\left(L,t\right)$ and $I\left(L,t\right)$
falls down exponentially with a time constant of about $T_{1}$ when
the incident beam is switched off. In these experiments, the signal
hardly exceeds the dark signal of the detector and the signal-to-noise
ratio is poor compared to that obtained by Kozlov \emph{et al.} and
in our experiments. The shape of the observed signals strangely resembles
that of the evolution of the fluorescent metastable level population
$N_{F}$ obtained in the incoherent bleaching model (Fig. \ref{fig:Figure5}).
It thus seems that the\emph{ light actually observed in} \cite{wi14}\emph{
is nothing but fluorescence light}. The fact that ``the delays of
the individual Fourier components {[}of the observed signal{]} are
independent of the modulation frequency\textquotedblright{} as noted
in \cite{wi14} simply reflects that, for moderate saturations, the
population $N_{F}(L,t)$ practically evolves as the voltage in an RC
circuit {[}see inset of Fig. \ref {fig:Figure5} and Eqs. (\ref{eq:equationVingtSept} and (\ref{eq:equationVingthuit}){]}.
Wisniewski\emph{ et al.} add ``the shape of the tail should be independent
of the modulation frequency of the pulse\textquotedblright . Equation. (\ref{eq:equationVingtSept})
shows that it is actually the case for $N_{F}(L,t)$ whatever the
saturation is. Otherwise said, all the experimental results reported
in \cite{wi14} are quite compatible with the incoherent bleaching
model insofar as the observed light seems to be fluorescence light
and not the light transmitted at the laser frequency.

\section{Summary and discussion\label{sec:Conclusion}}

Our article confirms the validity of the bleaching model to analyze
the slow transmission of light pulses through saturable absorbers.
Most experiments of so-called ``Slow light based on (by means of,
via) \emph{coherent} population oscillations\textquotedblright{} are
fully explained by the\emph{ incoherent} bleaching model. The very
expression of coherent population oscillations is misleading. Indeed
coherence in the optical sense generally plays no role in these experiments
and the population oscillations at the low frequency of the intensity
modulation are a trivial consequence of the equations coupling intensity
and populations in saturable absorbers \cite{gi65,se67,se70,se71,pi07,ma08}.
The population change being delayed with respect to the modulation
(Fig. \ref{fig:Figure1}), the attribution of the origin of the slow
transmission of the modulation to the population oscillations is questionable.
We finally remark that the concept of group velocity as defined for
pulses of coherent light does not apply to the broadband light considered
in most experiments. It is, however, possible to define a group delay
for the modulation {[}Eq. (\ref{eq:equationDixsept}){]} and even to
establish a generalized Kramers-Kronig relation between its phase
and its amplitude {[}Eq. (\ref{eq:equation8}){]}. Note that these
results are consistent with the CPO model as revised in \cite{pi07}
and that the equations obtained in the latter are only local expressions
of the integral equations derived in the incoherent bleaching model.
An important point is that the modulation group delay has an upper
limit equal to the medium response time $T_{1}$ no matter the medium
thickness and is often much shorter. See Fig. \ref{fig:Figure1} and \ref{fig:Figure2}. This situation contrasts with that encountered
in ``pure\textquotedblright{} slow light experiments performed with
coherent light where the time delays do not suffer such limitations
\cite{bo05}. As soundly remarked in \cite{se09}, it thus appears
that the slow transmission in saturable absorbers reflects ``slow
response'' of the medium rather than, strictly speaking, ``slow
light''.

The above analysis essentially concerns the transmission of light
pulses superimposed on a large background. For saturating pulses without
background, the range of validity of the CPO model, based on linearized
equations, is artificially extended in \cite{bi03,wi14} by invoking
a mechanism of self-pumping, one part of the pulse acting as a pump
and the remaining part as a probe. Without any theoretical justification,
this model gives pulse shapes that are qualitatively inconsistent
with those derived in the incoherent bleaching model \cite{se67,se70,ma08}
and actually observed for organic dyes \cite{kh67,he68}. The recent
claim of the failure of the incoherent bleaching model to explain
the pulse shapes observed in the reference case of ruby at room temperature
\cite{wi13} originates two complementary experiments with this material
\cite{ko14,wi14}. Incident square wave pulses were used in these
experiments, the main point being that, in the bleaching model, the
transmitted light should immediately vanish at the instant where the
incident light is switched off. The experiments were performed with
similar setups but, surprisingly enough, gave opposite results. The
first one \cite{ko14} validates the bleaching model whereas the following
one \cite{wi14} shows an exponential-like fall of the transmitted
pulse when the incident light is switched off. No explanation was
given in \cite{wi14} of this discrepancy. Our experiments and our
theoretical calculations of the transmitted pulse shape and of the
ruby fluorescent metastable state population bring some light on the
problem. Our experimental result (Fig. \ref{fig:Figure3}) not only
shows that the observed signal falls to 0 at the end of the square
but that its rise is also in agreement with that predicted by the
bleaching model (Fig. \ref{fig:Figure4}). The apparently opposite
result obtained in \cite{wi14} can be explained by examining the
theoretical evolution of the population of the fluorescent state (Fig. \ref{fig:Figure5})
that strangely resembles the observed signal in \cite{wi14}. We
believe that the latter is caused by the fluorescence and the incoherent
bleaching model is thus entirely validated.                    
   
\section*{}

	We are grateful to Professor E. Zhukov at Dortmund Technical University
who kindly provided us the ruby crystal rod used in our experiments.
This work has been partially supported by Ministry of Higher Education
and Research, Nord-Pas de Calais Regional Council and European Regional
Development Fund (ERDF) through the Contrat de Projets État-Région
(CPER) 2015\textendash 2020, as well as by the Agence Nationale de
la Recherche through the LABEX CEMPI project (ANR-11-LABX-0007).


\begin{thebibliography}{10}
\bibitem{re1}Throughout our paper, we use lower-case (upper-case)
omega to designate optical frequencies (low modulation frequencies).
\bibitem{re3}Note that the oscillation frequency involved in \cite{go02}
ranges in the microwave domain while only audio-frequencies take place
in the experiments on saturable absorbers.
\bibitem{re2} We incidentally remark that the experimental results
of \cite{bi03,bi04,ba05} have not been reanalyzed accordingly.
\end{thebibliography}
\end{document}